\pdfoutput=1
\documentclass[a4paper,american,citeautoscript,floatfix,pdftex,superscriptaddress,twoside,%
aps,pra,%
longbibliography,%
reprint,twocolumn,final
]{revtex4-1}%
\usepackage{amsfonts,amsmath,amssymb}
\usepackage{commath}
\usepackage[T1]{fontenc}
\usepackage{graphicx}%
\usepackage[utf8]{inputenc}
\usepackage{microtype}
\usepackage[obeyFinal,textsize=scriptsize]{todonotes}
\usepackage{xspace}
\usepackage{hyperref, hypernat}
\usepackage[todos]{CMImacros}

\graphicspath{{./figures/}} 
\hypersetup{citebordercolor=yellow,linkbordercolor=red,urlbordercolor=blue,hidelinks} %

\newcommand{\cfeldesy}{\affiliation{Center for Free-Electron Laser Science, Deutsches
      Elektronen-Synchrotron DESY, Notkestraße 85, 22607 Hamburg, Germany}}%
\newcommand{\uhhchem}{\affiliation{Department of Chemistry, Universität Hamburg,
      Martin-Luther-King-Platz 6, 20146 Hamburg, Germany}}%
\newcommand{\uhhcui}{\affiliation{The Hamburg Center for Ultrafast Imaging, Universität Hamburg,
      Luruper Chaussee 149, 22761 Hamburg, Germany}}%
\newcommand{\uhhphys}{\affiliation{Department of Physics, Universität Hamburg, Luruper Chaussee 149,
      22761 Hamburg, Germany}}%
\newcommand{\jkemail}{\email[]{jochen.kuepper@cfel.de}}%
\newcommand{\cmiweb}{\homepage{https://www.controlled-molecule-imaging.org}}%

\begin{document}
\onecolumngrid%
\title{Strong-field photoelectron momentum imaging of OCS \texorpdfstring{\\}{} at finely resolved
   incident intensities}%
\author{Joss Wiese}\cfeldesy\uhhchem%
\author{Jean-François Olivieri}\cfeldesy%
\author{Andrea Trabattoni}\cfeldesy\uhhcui%
\author{Sebastian Trippel}\cfeldesy\uhhcui%
\author{Jochen Küpper}\jkemail\cmiweb\cfeldesy\uhhchem\uhhcui\uhhphys
\begin{abstract}\noindent%
   Photoelectron momentum distributions from strong-field ionization of carbonyl sulfide with 800~nm
   central-wavelength laser pulses at various peak intensities from 4.6
   to~$13\times10^{13}$~\Wpcmcm were recorded
   and analyzed regarding resonant Rydberg states and photoelectron orbital angular momentum. The
   evaluation of the differentials of the momentum distributions with respect to the peak intensity
   highly suppressed the impact of focal volume averaging and allowed for the unambiguous
   recognition of Freeman resonances. As a result, previously made assignments of photoelectron
   lines could be reassigned. An earlier reported empirical rule, which relates the initial state's
   orbital momentum and the minimum photon expense to ionize an ac Stark shifted atomic system to
   the observable dominant photoelectron orbital momentum, was confirmed for the molecular target.
\end{abstract}
\maketitle

\section{Introduction}
\label{sec:introduction}
Studies in strong-field physics aim at the understanding and control of the electron wave packet
emitted by an atomic or molecular target following the exposure to intense radiation. Vast
theoretical and experimental efforts are focused on, \eg, the determination of intensity-dependent
ionization probabilities~\cite{Hankin:PRA64:013405, Hart:PRA89:1393} and cut-off energies for direct
and rescattered electrons~\cite{Hart:PRA89:1393}, the recognition of Freeman
resonances~\cite{Shao:PRA:2008, Yu:JPB50:235602, Li:PRA92:1945}, the observation of
channel-switching and -closing effects~\cite{Schyja:PRA57:3692, Marchenko:JPB43:095601}, and the
analysis of the photoelectron orbital angular momentum~\cite{Marchenko:JPB43:095601, Li:PRA92:1945}
in the multi-photon regime as well as the imaging of the initial state's electron density
distribution~\cite{Holmegaard:NatPhys6:428, Dimitrovski:JPB48:245601, Maurer:PRL109:123001}. In
general for strong-field ionization, the electron wave packet's initial distribution in phase space
-- and thereby its subsequent dynamics in the field -- is strongly shaped by the intensity of the
electric field~\cite{Ivanov:JMODOPT52:165}. Within the framework of multi-photon ionization the
intensity-dependent ac Stark shift alters the energies of initial and final target states as well as
of any intermediate state that is resonantly passed through. In addition, the continuum the electron
is born into is raised by the intensity-dependent ponderomotive energy. From the perspective of
tunneling ionization the intensity dictates the shape of the potential barrier to be traversed in
order to reach the continuum~\cite{Ammosov:SVJETP64:1191}. As a result, the outgoing electron wave
packet is substantially dependent not only on the target system, but also on the intensity of the
driving field.

Resolving the incident intensity in a strong-field experiment is as fundamental as knowing the
incident photon energy in optical spectroscopy. Especially for the comparison of experimental data
with predictions from strong-field ionization models intensity selectivity is highly beneficial,
since intensity integration does not blur the results. However, in experimental investigations
employing typical laser-focus geometries the recorded target response results from intensities
ranging from 0 to the peak intensity. As a consequence, the target information encoded in the
experimental observables is often highly obscured by integration over all incident intensities.
Various schemes were reported that allow the investigator to overcome this issue and enable access
to the target response in an intensity-resolved fashion. An intensity-selective scanning method,
which relies on the extraction of charged particles created within a restricted slice of the
longitudinal intensity distribution -- implemented through an aperture -- in conjunction with an
intensity deconvolution step in post-processing was introduced~\cite{Walker:PRA57:R701} and later
further refined~\cite{Hart:PRA89:1393}. For \emph{cw} laser beams the acquisition of
intensity-difference spectra was described~\cite{Wang:OL30:664}. The approach proposed is based on
the evaluation of the derivative of the target response with respect to the peak intensity. Through
analytical inversion by means of a power-series expansion in intensity, namely multi-photon
expansion technique, the retrieval of distinct-intensity responses from a set of intensity-averaged
measurements at different peak intensities was demonstrated~\cite{Strohaber:PRA82:013403}. While the
intensity-selective scanning and the multi-photon expansion technique employ rather complex
post-processing, they essentially eliminate the effect of focal intensity averaging depending on the
peak-intensity step size of the data set. Although through the acquisition of intensity-difference
spectra, using the most common focus geometry, the impact of intensity averaging is only highly
suppressed yet not completely removed, it represents the by far simplest approach of the three.

In this manuscript, the advantages of intensity-difference spectra are studied for pulsed laser
beams and utilized to investigate the intensity-dependent photoelectron momentum distributions from
strong-field ionization of carbonyl sulfide (OCS) molecules in the intermediate regime between
multi-photon and tunneling ionization.

\section{Intensity-difference signals}
\label{sec:diprobing}
The differential volume occupied by a distinct intensity in a \emph{cw} laser beam can be readily
described~\cite{Wang:OL30:664}. This focal intensity distribution is referred to as one
implementation of a 2D configuration. In the following, the corresponding case of a pulsed laser
beam will be investigated.

\begin{figure}
   \includegraphics{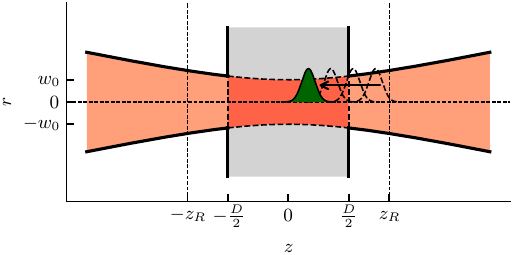}%
   \caption{Schematic view of the focal geometry: The laser beam (orange) intersects the gas-target
      beam (gray), which is assumed to possess a uniform density distribution. If the diameter of
      the gas-target beam, $D$, is smaller than twice the Rayleigh length of the focus, $z_R$, the
      effective focal volume can be approximately described with a constant radial standard
      deviation, $\sigma_r=w_0/2$. The spatial equivalent of the laser pulse's temporal distribution
      for a distinct time is a Gaussian profile (green). Since neither the longitudinal nor the
      radial intensity distribution change upon the pulse's passage through the focal volume, the
      whole space- and time-integrated laser-target interaction can be described by a spheroidal
      intensity function, see \eqref{eq:idist}.}
   \label{fig:focus}
\end{figure}

The intensity profile of an effective laser focus in a typical gas-phase experiment is assumed to
have the following geometric properties: in space it is rotation-symmetric about the wave vector and
its distribution along the propagation direction is given by a rectangular function with the width
of the diameter of the skimmed gas-target beam $D$, relying on $D$ to be smaller than twice the
focus's Rayleigh length $z_R$. \autoref{fig:focus} provides a schematic illustration of the focus
geometry. The transverse spatial and the temporal profile follow normal distributions with standard
deviations $\sigma_r$ and $\sigma_t$. For a distinct time during the laser pulse's propagation
through the focus volume the three-dimensional intensity distribution can be written as
\begin{equation}
   I = I_0 \cdot \exp\left(-\frac{r^2}{2\sigma_r^2}\right)
   \cdot \exp\left(-\frac{z^2}{2\sigma_z^2}\right).
   \label{eq:idist}
\end{equation}
Here, $I_0$ is the temporal peak intensity and $\sigma_z=c\sigma_t$ is the spatial equivalent of the
temporal standard deviation $\sigma_t$. If $D \gg \sigma_z$, the full intensity spectrum, \ie, for
all possible longitudinal positions of the pulse, will be quasi-proportional to the intensity
spectrum at a single position far away from the longitudinal borders given by the edges of the
gas-target beam. Thus, the intensity distribution displayed above can be used to deduce the
intensity spectrum of the complete space- and time-integrated focus interaction. In the following,
a focus with such an intensity distribution is referred to as 3D configuration.
After rearrangement of \eqref{eq:idist} to
\begin{equation}
   \frac{r^2}{\sigma_r^2} + \frac{z^2}{\sigma_z^2} = 2 \, \ln\frac{I_0}{I} ,
\end{equation}
it is obvious that points of equal intensity are located on the surface of a spheroid. Accordingly,
the volume function is
\begin{equation}
   V_\text{3D}(I) = \frac{4\pi}{3} \sigma_r^2 \sigma_z \left(2\ln\frac{I_0}{I}
   \right)^{3/2}
   \label{eq:V}
\end{equation}
with the differential volume $-({\partial{V_\text{3D}}}/{\partial{I}})\dif{I}$ that is occupied by
the iso-intensity shell $\dif{I}$ around $I$ following
\begin{equation}
   -\frac{\partial V_\text{3D}}{\partial I} \dif{I} \propto
   \frac{\sqrt{\text{ln}\frac{I_0}{I}}}{I} \dif{I} .
   \label{eq:dV}
\end{equation}
This provides a measure for the weight of any intensity between 0 and $I_0$ in the laser--target
interaction. It complements the previous elaborations~\cite{Wang:OL30:664} by the additional
consideration of the laser pulse's temporal intensity profile. Note that the description of the
volume function $V_\text{3D}$ will only be as simple as \eqref{eq:V} if the radial intensity
distribution is independent of $z$, \ie, $D<2z_R$.

\begin{figure}
   \includegraphics{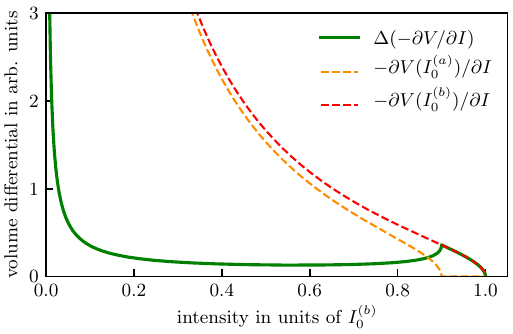}
   \caption{Intensity-dependent volume differentials for a space- and time-integrated laser--target
      interaction in a 3D configuration for two peak intensities $I_0^{(a)}/I_0^{(b)}=0.9$
      (dashed lines). The difference between the two volume differentials (solid line) exhibits
      highly suppressed intensity contributions for $I<I_0^{(a)}$.}
   \label{fig:ispec}
\end{figure}
\autoref{fig:ispec} shows the resulting intensity spectra for two peak intensities
$I_0^{(a)}/I_0^{(b)}=0.9$ along with the corresponding difference spectrum. Evidently, in the
difference spectrum the contributions from intensities $<I_0^{(a)}$ are highly suppressed, albeit
not completely eliminated. The pole point at $I=0$ does not affect the determination of any
intensity-dependent quantity, because the corresponding signal scales with
$\Delta(-\partial V_\text{3D}/\partial I)\cdot{}I^n$, $n>0$. Here, $n$ would be an integer for a
pure single-channel multiphoton process but does rather represent an intensity-dependent quantity
for a realistic strong-field interaction.
\begin{figure*}
   \includegraphics{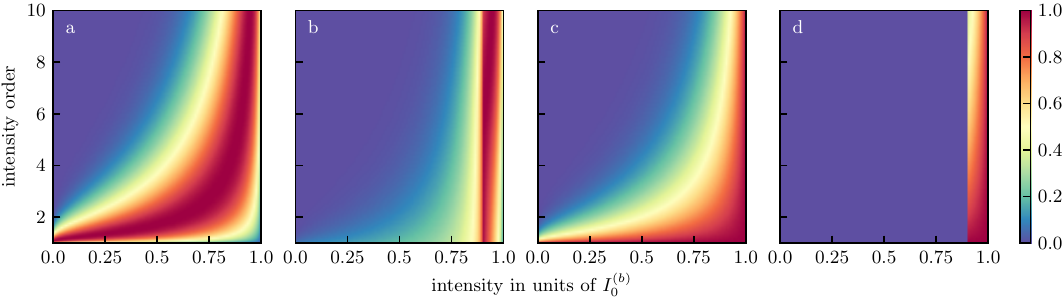}
   \caption{Normalized signals as a function of intensity and intensity order $n$ for different
   volume differentials: a) full volume in 3D configuration, b) differential volume in 3D
   configuration, c) full volume in 2D configuration, d) differential volume in 2D configuration.
   These provide a measure for the relative contribution of a distinct intensity to an overall
   signal, that scales with intensity to the power of a fixed exponent. For the differential-volume
   signals a ratio of $Q=I_0^{(a)}/I_0^{(b)}=0.9$ was used, which represents the lowest $Q$ value
   employed in experiment. All signal maps depicted were row-wise normalized to their maxima. For
   the 2D configuration case the volume differential follows
   $-{\partial V_\text{2D}/\partial I} \propto -{1/I}$~\cite{Wang:OL30:664}.}
   \label{fig:sdv}
\end{figure*}
The full-volume and exemplary difference signals are shown in \autoref{fig:sdv} for both, the 2D and
the 3D configuration, as a function of intensity and intensity order $n$. Note that for a focal
interaction with intensities normally distributed only along two dimensions (2D configuration), \ie,
either a spatial laser sheet with a Gaussian temporal profile or a two-dimensional spatial profile
with a flat-top temporal profile, the corresponding difference intensity spectrum carries non-zero
contributions only for $I_0^{(a)}\leq{}I<I_0^{(b)}$.
\begin{figure}[b]
   \includegraphics{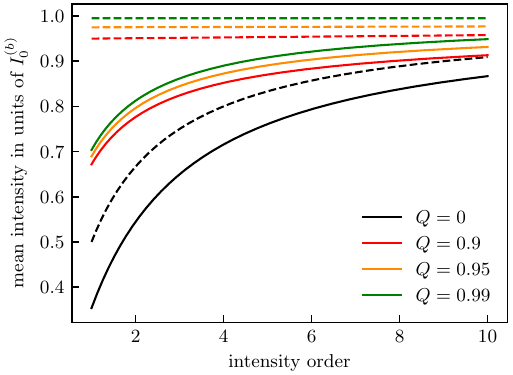}
   \caption{Mean intensity as a function of the intensity order $n$ for different
      intensity-difference signals with peak-intensity ratios $Q = I_0^{(a)}/I_0^{(b)}$ for the 3D
      (solid lines) and 2D configurations (dashed lines). The black lines correspond to signals at
      full volume integration, $Q=0$, \ie, evaluated at a single peak intensity. $Q$ values of $0.9$
      and $0.99$ represent the lower and upper limits of peak-intensity ratios used in our
      experiments, \emph{vide infra}.}
   \label{fig:im}
\end{figure}
\autoref{fig:im} displays the mean intensities for various intensity-difference signals in
dependence of the intensity order. In general, the use of a 2D focus configuration gives rise to
mean intensities that are closer to the peak intensity compared to the 3D case. The black lines
illustrate the change of the mean intensity with increasing intensity order for measurements at a
single peak intensity. Especially for low intensity orders the corresponding mean intensities are
far below the (upper) peak intensity. In contrast, in an intensity-difference setup the situation is
much better, further improving with rising peak-intensity ratios $Q=I_0^{(a)}/I_0^{(b)}$. The green
graphs, $Q=0.99$, provide insight into the achievable mean intensity for $Q\to1$: for the 2D
configuration the mean intensity reflects, almost independently of the intensity order, the actual
peak intensity. However, for the 3D case there is an appreciable deviation, especially for signals
that arise at low intensity orders. Furthermore, as $Q$ increases, the fractional signal that is 
kept with respect to the full signal at peak intensity becomes smaller.
\begin{figure}[b]
   \includegraphics{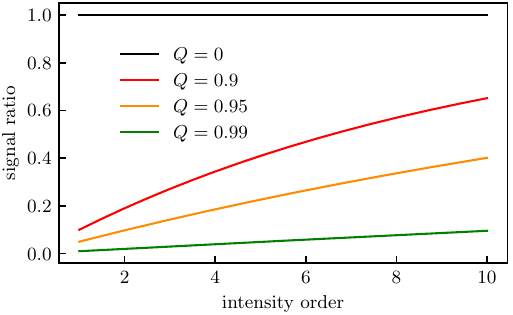}
   \caption{Fraction of the intensity-difference signal of the overall signal at peak intensity. 2D
      and 3D configuration show identical behavior.}
   \label{fig:srat}
\end{figure}
This behavior is depicted in \autoref{fig:srat} and is identical for 2D and 3D configurations. In
actual experiments one needs to find a compromise between intensity resolution and remaining signal.

In the following paragraph intensity-integrated observables and their derivatives with respect to
peak intensity are elucidated. Any $M$-dimensional observable $S(I_0)$, \eg, a two-dimensional
projection of photoelectron momenta, that results from the full laser--target interaction at peak
intensity $I_0$, obeys the relation
\begin{equation}
   S(I_0) \propto \int_0^{I_0} \Omega(I) s(I) \left(-\frac{\partial V}{\partial I} \right)
   \dif{I} ,
   \label{eq:S}
\end{equation}
with a normalized $M$-dimensional distribution $s(I)$ that is created at the distinct intensity $I$
with probability $\Omega(I)$ and the volume $-(\partial V/\partial I) \dif{I}$ occupied by the
iso-intensity shell d$I$. If $-(\partial V/\partial I) \dif{I}$ is independent of $I_0$ (2D
configuration), the evaluation of $\dif{S}/\dif{I_0}$ enables the direct determination of
$\Omega(I) s(I)$ by elimination of the integral sign in \eqref{eq:S}. This is not viable for the 3D
configuration, since the volume differential is a function of $I_0$. But also for this case
$\dif{S}/\dif{I_0}$ is proportional to an effective distribution $\tilde{s}$, that arises
from a restricted intensity spectrum as depicted in \autoref{fig:ispec} and \ref{fig:sdv}. Thus, the
acquisition of $\dif{S}/\dif{I_0}$, or its discrete equivalent
$\Delta S/\Delta I_0$, allows for the investigation of intensity-dependent changes in the
signal distribution, while intensity-dependent occurrence probabilities, $\Omega(I)$, are not
accessible.

Although a 2D focus configuration would enable the determination of intensity-dependent ionization
rates as well as momentum distributions without contamination from intensities $I < I_0^{(a)}$, the
experimental realization of such a setup suited for intensities appropriate for strong-field
ionization studies proves to be challenging. However, the evaluation of $\Delta{S}/\Delta{I_0}$ in a
3D configuration is easy to implement and provides the investigator with incident-intensity resolved
signal distributions $\tilde{s}$. This enables analyses of the intensity-dependence of an
$M$-dimensional signal, while the information about the integral signal remains obscured.

\section{Experimental setup}
\label{sec:experimental}
The experimental setup was described elsewhere~\cite{Trippel:MP111:1738, Chang:IRPC34:557}. In
brief, a cold molecular beam was created by supersonically expanding OCS seeded in 85~bar of helium
through an Even-Lavie valve~\cite{Even:JCP112:8068} operated at a repetition rate of 250~Hz. The
skimmed molecular beam was intersected in the center of a velocity-map imaging spectrometer by a
laser beam from a mode-locked Ti:sapphire system, which delivered Fourier-limited pulses at 800~nm
central wavelength and 35~fs pulse duration (full width at half maximum) at a repetition rate of
1~kHz. Employing a 500~mm focal length biconvex lens resulted in a laser focus with a beam waist of
44~\um and a Rayleigh length of $z_R=7.5$~mm. The diameter of the molecular beam was found to be
$D = 2.2$~mm and $2z_R>D$ provides justification for treating the longitudinal intensity
distribution of the effective focus as a rectangular function in the previous section. The pulse
energy reaching the interaction region was defined using a half-wave plate and a thin-film
polarizer. Ionization of OCS could be studied employing peak intensities of
$4.6\text{--}13\times10^{13}~\Wpcmcm$, respectively. Due to the charge-dissipation limit of the
in-vacuum detector components measurements at $I_0>8\times10^{13}~\Wpcmcm$ were conducted at a
reduced sample density -- namely $1/3$ of the density employed for measurements at
$I_0\leq8\times10^{13}~\Wpcmcm$. This was achieved by delaying the laser pulses with respect to the
maximum of the molecular beam's temporal profile. Sample densities of $1\times10^8$~cm$^{-3}$ and
$1/3\times10^8$~cm$^{-3}$ were utilized, respectively. Peak intensities as determined by a
combination of beam-profiling, auto-correlation, and measurement of the pulse energy were found to
be too low by $20~\%$ compared to the intensity-dependent kinetic energy shift of distinct
above-threshold ionization (ATI) peaks, see \eqref{eq:Tfin}. In combination with the known energy
spacing between adjacent ATI peaks, namely the photon energy $h\nu$, this approach allowed for an
intrinsic calibration of both, the photoelectron momentum and the laser intensity. Two-dimensional
projections of photoelectron momenta were recorded by mapping the electrons onto a
position-sensitive detector, consisting of a microchannel plate (MCP), a phosphor screen and a
high-frame-rate camera (Optronis~CL600), by means of a velocity map imaging spectrometer (VMIS).
This detection scheme allowed for counting and two-dimensional momentum-stamping of single
electrons. The momentum resolution achieved -- given by the voltage applied to the VMIS electrodes
and the pixel size of the camera -- was $5.6\times10^{-3}$~atomic units (electron velocity of
12~km/s). Momentum slices of the full three-dimensional momentum distributions $S(I_0)$ were
recovered through inverse Abel-transformations employing the ``Onion-Bordas'' onion-peeling
algorithm~\cite{Bordas:RSI67:2257, Rallis:RSI85:113105} as implemented in the \texttt{PyAbel}
software package~\cite{Hickstein:pyabel}.

\section{Results and Discussion}
\label{sec:resdis}
\begin{figure} 
   \includegraphics{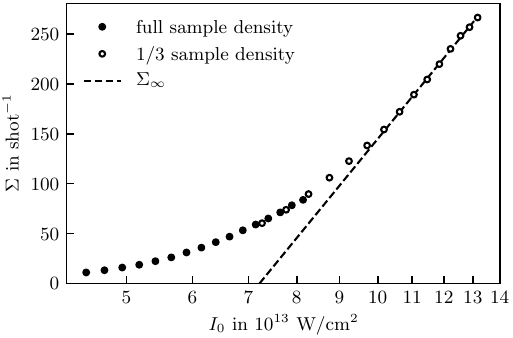}
   \caption{Experimental electron count rates versus the peak intensity on a semi-logarithmic scale
      (dots). Open dots represent measurements conducted at 1/3 of the maximum sample density
      rescaled accordingly to match those recorded at full density. The asymptotic behavior beyond
      the saturation intensity is depicted by the dashed line.}
   \label{fig:rho}
\end{figure}
\begin{figure*} 
   \centering%
   \includegraphics{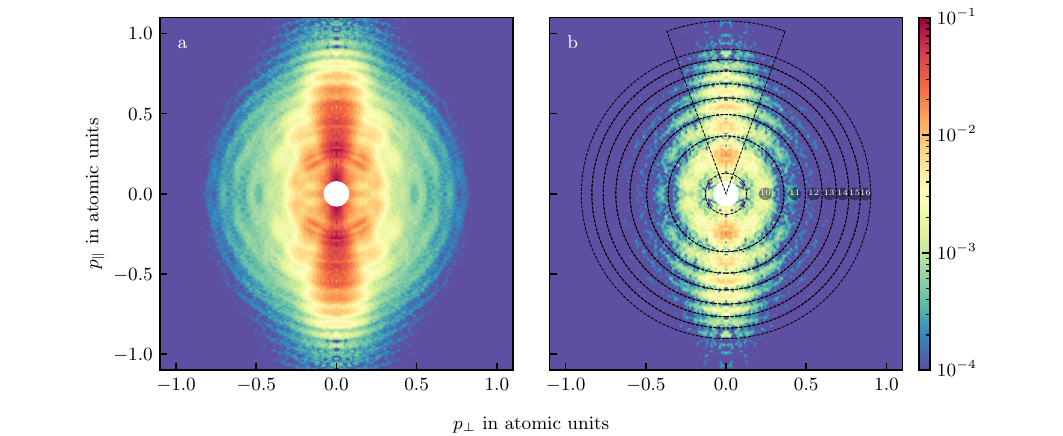}
   \caption{Photoelectron momentum slices a) for the full-intensity distribution $S$ at a peak
      intensity of $5.7\times10^{13}~\Wpcmcm$ and b) for the differential-intensity distribution
      $S'$, which is dominated by intensities of $5.4\text{--}5.7\times10^{13}~\Wpcmcm$. The
      black-line wedge sketches the polar angles used to evaluate the intensity-dependent radial
      distributions shown in \autoref{fig:radfull} and \ref{fig:raddiff}; the black circles
      separate contributions from distinct photon numbers $N$ and correspond to kinetic energies
      $T(N\pm0.5,\tilde{I}_0)$ as obtained from \eqref{eq:Tfin}.}
   \label{fig:pmaps}
\end{figure*}
\subsection{Integral photoelectron yield}
Extensive strong-field ionization studies~\cite{Hankin:PRA64:013405} employing various laser peak
intensities demonstrated that the asymptotic slope of an integral ionization signal $\Sigma(I_0)$
with respect to $\ln{I_0}$, arising from the full intensity spectrum of the laser–target
interaction at peak intensity $I_0$, follows
\begin{equation}
   \lim_{I_0\to\infty} \left( \frac{\dif{\Sigma}}{\dif\ln{I_0}} \right) = 2 \pi \alpha \sigma_r^2 D \varrho \, .
\end{equation}
$\alpha$ is the instrument sensitivity, $\sigma_r$ the standard deviation of the transverse
intensity distribution, $D$ the length of the focus volume and $\varrho$ the sample density. The
only requirements for the validity of this equation are the predominance of single ionization and
the cylindrical symmetry of the spatial focus geometry. Due to the onset of saturation, starting at
intensity $I_\text{sat}$, the slope of $\Sigma(\ln{I_0})$ converges to a constant value for
increasing peak intensity. Within the notation of this manuscript $\Sigma(I_0)$ equals the
momentum-space integral of the distribution $S(I_0)$, which for the present case is just the total
number of electrons detected. \autoref{fig:rho} shows the resulting values of $\Sigma$ plotted
versus $I_0$ on a semi-logarithmic scale. With increasing values of $\text{ln}I_0$ the slope of the
total electron yield converges to a constant value. In the following analysis, this behavior is
used as implicit evidence for a negligible contribution from multiple ionization. The asymptotic
slope and the saturation onset were deduced by modeling a line through the highest peak-intensity
points providing a saturation onset of $I_\text{sat}=7.2\times10^{13}~\Wpcmcm$. The saturation
intensity is a function of both, the pulse duration and the orientation-dependent ionization
probability~\cite{Hankin:PRA64:013405}. Using the MCP's open area ratio of 0.64 as instrument
sensitivity $\alpha$, a transverse focal standard deviation of $\sigma_r=22~\um$ (beam waist of
44~\um) and a molecular beam diameter of $D=2.2$~mm resulted in a sample density of
$\varrho=1\times10^{8}~\Wpcmcmcm$. Since the intensity could be intrinsically calibrated from the
ponderomotive shift of the ATI peaks (see next subsection), the deduced sample density is as
accurate as $\alpha \sigma_r^2 D$.

\subsection{Radial and angular distributions}
Due to their rotational symmetry the three-dimensional momentum distributions $S(I_0)$ can be
completely described by two-dimensional slices of momenta parallel ($p_\parallel$) and perpendicular
($p_\bot$) to the polarization axis. In the following, the slope of $S$ with respect to $I_0$,
\begin{equation}
   S'(\tilde{I}_0) = \frac{\Delta S}{I_0^{(b)} - I_0^{(a)}} \, ,
\end{equation}
is analyzed regarding its radial and angular distributions. Here, the mean intensity $\tilde{I}_0$
corresponds to the arithmetic mean of every two peak intensities $I_0^{(b)}>I_0^{(a)}$ that were
used to measure $S$. Since $S'(\tilde{I}_0)$ arose from a more narrow distribution of incident
intensities than $S(I_0)$, it allows for the detailed investigation of intensity dependent effects.

Both $S(I_0)$ and $S'(\tilde{I}_0)$ exhibit rich angular and radial structure. \autoref{fig:pmaps}
displays a typical momentum slice for the two cases, here for intensities around
$5.5\times10^{13}~\Wpcmcm$. Both momentum distributions show broad radial patterns that repeat with
integer multiples of the photon energy -- referred to as ATI structure --, sharp radial features on
top of the ATI structures -- referred to as Freeman resonances~\cite{Freeman:PRL59:1092} -- and a
carpet-like pattern perpendicular to the polarization axis~\cite{Korneev:PRL108:223601}.

\subsubsection{Radial distributions}
\begin{figure}
   \includegraphics{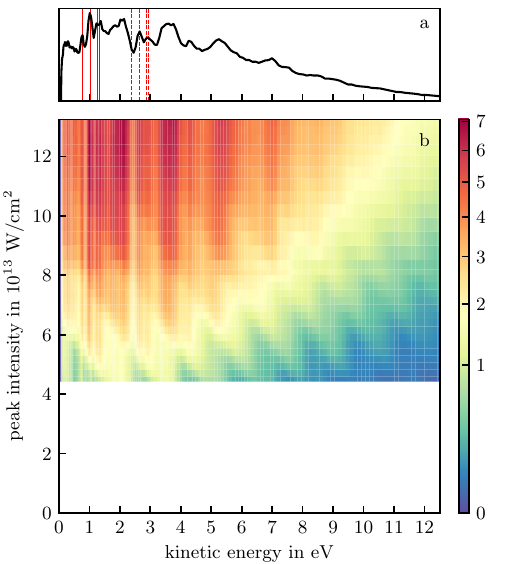}
   \caption{Radial distributions of $S(I_0)$ were evaluated for polar angles in
      $[\degree{-20}, \degree{+20}]$ as sketched in \autoref{fig:pmaps}b: a)
      intensity-integrated radial distribution carrying imprints of Freeman resonances with
      principal quantum numbers $n=4,5,7,8$ indicated for $k=1$ (solid red lines) and $k=2$ (dashed
      red lines) as obtained through \eqref{eq:TF} and b) peak-intensity-dependent radial
      distributions. Values of $S(I_0)$ are mapped onto a square-root color scale.}
   \label{fig:radfull}
\end{figure}
In \autoref{fig:radfull} the peak-intensity dependent radial distributions for polar angles between
\degree{-20} and \degree{+20} with respect to the polarization axis are illustrated for $S(I_0)$.
There are two kinds of features to be noted: (\emph{i}) broad peaks with a spacing of $h\nu$ that
experience a shift in kinetic energy proportional to the negative value of the peak intensity and
(\emph{ii}) sharp peaks at constant kinetic energies. The radial distributions of $S(I_0)$ are
dominated by feature (\emph{ii}).
\begin{figure}
   \includegraphics{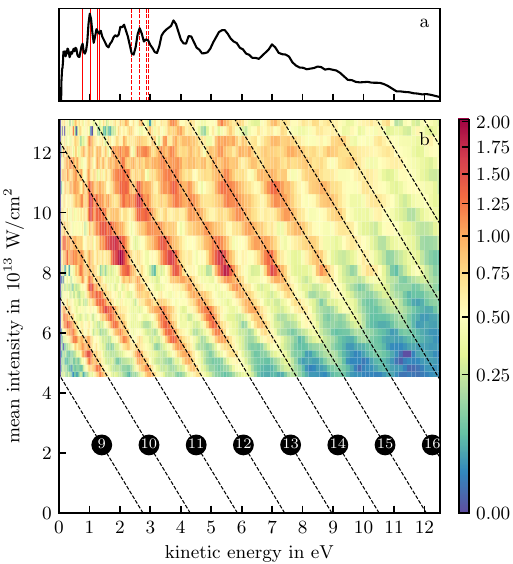}
   \caption{Radial distributions of $S'(\tilde{I}_0)$ were evaluated for polar angles in
      $[-20^\circ, +20^\circ]$ as sketched in \autoref{fig:pmaps}b: a) intensity-integrated
      radial distribution and b) intensity-dependent radial distributions. Broken black
      lines in b) depict ponderomotively shifted final kinetic energies as obtained through
      \eqref{eq:Tfin} for the given numbers of photons. Values of $S'(\tilde{I}_0)$ are mapped onto
      a square-root color scale.}
   \label{fig:raddiff}
\end{figure}
\autoref{fig:raddiff} displays the equivalent radial distributions for $S'(\tilde{I}_0)$, which
exhibit predominant contributions from feature (\emph{i}). Note that there is a change in sample
density and intensity bin size at $8.3\times10^{13}~\Wpcmcm$, \cf \autoref{fig:rho}.

In the following, feature (\emph{i}), which corresponds to non-resonant above-threshold ionization,
is analyzed employing the radial distributions of $S'(\tilde{I}_0)$. Dashed black lines in
\autoref{fig:raddiff} correspond to final kinetic energies~\cite{Freeman:JPB24:325}
\begin{equation}
   T(N, \tilde{I}_0) = Nh\nu - \Ip{}^{(0)} - \Up{}(\tilde{I}_0) \label{eq:Tfin} \, ,
\end{equation}
with the number of absorbed photons $N$, the photon energy $h\nu = 1.55$~eV, the field-free
ionization potential of OCS, $\Ip{}^{(0)} = 11.19$~eV~\cite{NIST:webbook:2017}, and the
ponderomotive potential $\Up{}=\tilde{I}_0/4\omega^2$ (for atomic units). Clearly, \eqref{eq:Tfin}
describes the kinetic energies of the radial local maxima in \autoref{fig:raddiff} quite well
except for minor deviations towards lower kinetic energies at $T<3$~eV, which are ascribed to the
Coulomb interaction between photoelectron and cation. Thus, the total ionization energy appears to
be dictated by the ponderomotive continuum shift. For the intensity range of the current study
the relative Stark shift between the neutral and the cationic species is dominated by the
permanent-dipole interaction~\cite{Holmegaard:NatPhys6:428} and, hence, equals zero in average over
the isotropic distribution of molecular orientations.

Within the peak-intensity range investigated all radial distributions of $S(I_0)$ carry series of
sharp peaks, feature (\emph{ii}), that are also reproduced periodically with the photon energy
$h\nu$. Close inspection of \autoref{fig:raddiff} reveals that the kinetic energies of those peak
series are independent of the incident intensity, \ie, they appear as vertical lines imprinted onto
the non-resonant ATI pattern, feature (\emph{i}). Based on this
intensity-independence and the energy spacing between individual series, these peaks can
unambiguously be assigned to Freeman resonances~\cite{Freeman:PRL59:1092}, \ie, photoelectron
emission from resonance-enhanced multiphoton ionization through Rydberg states. Since those Rydberg
states experience the same ac Stark shift as their corresponding continua, namely the ponderomotive
energy, the resulting final kinetic energies $T_\text{F}$ are independent of the incident
intensity~\cite{Freeman:JPB24:325,OBrian:PRA49:R649}:
\begin{equation}
   T_{\text{F}}(k, n) = kh\nu - \frac{R_\infty}{\left( n + 0.25 \right)^2} \label{eq:TF} \, ,
\end{equation}
with the number of photons $k$ that ionize a resonant Rydberg state with principal quantum number
$n$ and the Rydberg constant $R_\infty$. The subtrahend reflects the energy difference between the
Rydberg state and its asymptotic unbound state. Resonant transitions for $n$ = 4, 5, 7 could be
observed together with a combination band for $n$ = 8--10, with $k$ = 1--3. These assignments
agree with previous single-photon absorption studies on OCS~\cite{Tanaka:JCP32:1205,
Sunanda:JQuantSpec113:58, CossartMagos:JCP119:3219}. In the present study individual angular
momentum states could not be resolved: at a kinetic energy of 1.2~eV, where the $k=1$ Freeman
manifold occurred, an energy resolution of 0.05~eV was achieved, which is roughly ten times larger
than what would be required to distinguish between different $l$ states.

Previous results on OCS for comparable peak intensities of $2.8$--$9.4\times10^{13}~\Wpcmcm$,
analyzed without the intensity-difference spectra, lacked information about the incident
ponderomotive shift and, hence, assigned intensity-independent photoelectron lines to
valence--valence transitions~\cite{Yu:JPB50:235602}. Three sharp peaks in the photoelectron kinetic
energy spectra, which follow $T_{1,3}=0.73\text{~eV}+\tilde{k}h\nu$, with $\tilde{k}=(0,1)$, and
$T_2 = 1.17\text{~eV}$, were attributed to the valence transitions
$^1\Delta\leftarrow\tilde{X}{^1\Sigma^+}$ and $^1\Pi\leftarrow\tilde{X}{^1\Sigma^+}$. The
corresponding excited valence states have field-free vertical ionization energies of roughly 5.8~eV
($^1\Delta$) and 3.9~eV ($^1\Pi$)~\cite{Sunanda:JQuantSpec113:58}. Both, the $^1\Delta$ and the
$^1\Pi$ state, were considered to experience ac Stark shifts equal to the ponderomotive potential,
resulting in intensity-independent kinetic energy releases~\cite{Yu:JPB50:235602}. Since \Up{} is
the kinetic energy imparted to an unbound electron in an oscillating electric field, it represents
the asymptotic ac Stark shift of an electronic state at the onset of the continuum. However,
treating the $^1\Delta$ and $^1\Pi$ states as quasi-unbound states is in conflict with their large
binding energies. Thus, in our present investigation the very same three peaks are, instead,
assigned to ionization of resonant Rydberg states leading to final kinetic energies of
$T_{1,3}=T_\text{F}(k=(1,2),n=4)$ and $T_2=T_\text{F}(k=1,n=7)$.

While $S'(\tilde{I}_0)$ mainly carries contributions from non-resonant above-threshold ionization,
$S(I_0)$ is dominated by resonance-enhanced multi-photon ionization. Through averaging over all
intensities present in the focal volume the ponderomotively shifted non-resonant contributions get
washed out in the final momentum distribution $S(I_0)$, while resonant contributions with constant
final momentum, \eg, Freeman resonances, gain intensity~\cite{Rudenko:JPB37:L407}. By inspection of
$S(I_0)$ only, the contributions from resonant ionization pathways are highly overestimated and
final kinetic energies cannot be related to distinct numbers of absorbed photons, because the
ponderomotive continuum shift \Up{} cannot be resolved. In the momentum distributions $S(I_0)$ the
closing of ionization channels with increasing peak intensities is highly obscured, because
contributions from all intensities between 0 and $I_0$ are contained. Thus, even if a channel is
already closed at the peak intensity, $S(I_0)$ will embody all contributions from $I < I_0$ with
even larger weight, \ie, larger differential volume, at which the channel is still open. In
principle, $S(I_0)$ contains just more and more open ionization channels with increasing peak
intensity. Nevertheless, experimental evidence for channel-closing was reported for xenon at fully
focal-volume averaged momentum distributions~\cite{Schyja:PRA57:3692, Li:PRA92:1945}. In the present
study, three channel-closings could be clearly observed, visible in \autoref{fig:raddiff} at
intensities when the ATI peaks assume zero kinetic energy.

\subsubsection{Angular distributions}
In the following, the non-resonant ATI channels for ionization with a minimum number of photons are
analyzed with respect to their photoelectron angular distributions. These channels give rise to the
lowest radial layers of the carpet-like pattern~\cite{Korneev:PRL108:223601} visible in
\autoref{fig:pmaps}a. The minimum number of photons needed to ionize the target is a staircase
function proceeding upward with increasing intensity. Accordingly, $S'(\tilde{I}_0)$ was sectioned
by the channel-closing intensities, which were obtained by evaluating the roots of \eqref{eq:Tfin}.
Within the intensity range investigated the closing of the $9,10,11,12$ photon channels were
encountered at intensities of $4.6,7.2,9.8,12.4\times10^{13}~\Wpcmcm$, respectively.
\begin{figure}
   \centering%
   \includegraphics{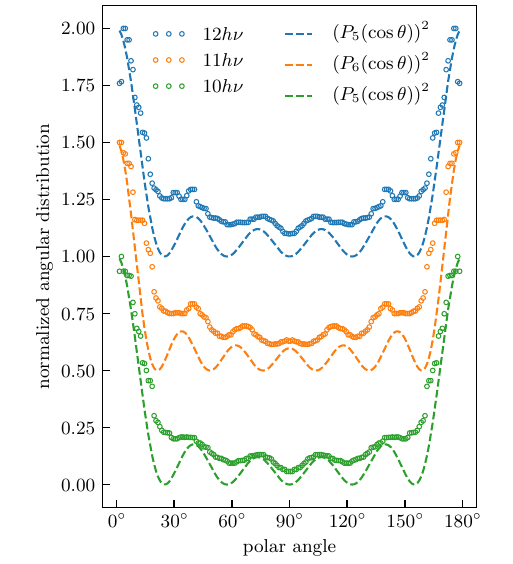}
   \caption{Angular distributions of the lowest order non-vanishing ATI channels for the minimum
      numbers of photons of $10,11,12$ required to ionize the OCS molecule. The three distributions
      were retrieved within the intensity intervals
      $[4.6,7.2[,[7.2,9.8[,[9.8,12.4[\times10^{13}~\Wpcmcm$, which represent the channel-closing
      intensities obtained by setting $T(N,\tilde{I}_0)=0$ in \eqref{eq:Tfin}. Experimental data
      are shown by circles, calculated traces of the squares of Legendre polynomials
      $P_l(\cos\theta)$ of order $l$ with the polar angle $\theta$ to the laser polarization
      axis are shown by dashed lines. The distributions are plotted with an arbitrary vertical
      offset for clarity.}
   \label{fig:angdist}
\end{figure}
In \autoref{fig:angdist}, circles depict the resulting angular distributions for ionization by
$10,11,12$ photons, each representing the lowest order non-vanishing ATI channel in its
corresponding intensity regime. The angular distributions associated with the absorption of
$10,11,12$ photons exhibit $5,6,5$ local minima, respectively.

As shown by a theoretical study of the strong-field ionization of atomic
hydrogen~\cite{Arbo:PRL96:143003} and by experimental investigations of xenon~\cite{Li:PRA92:1945},
the dominant orbital angular momentum quantum number $l$ of the emitted electrons can be deduced
from fitting the angular distributions by squared Legendre polynomials
$\left(P_l(\cos\theta)\right)^2$ of order $l$. According to the number of local minima the
absorption of $N=10,11,12$ photons gives rise to dominant orbital angular momentum quantum numbers
of $l=5,6,5$. Best matches of $\left(P_l(\cos\theta)\right)^2$ to the experimental angular
distributions are shown as dashed lines in \autoref{fig:angdist}.

An empirical rule was given that predicts the dominant photoelectron orbital angular momentum
quantum number $l$ based only on the minimum number of photons to reach the continuum and the
orbital quantum number $l_0$ of the initial bound electronic state~\cite{Chen:PRA74:053405}.
For intensities implying minimum numbers of $10,11,12$ photons needed to ionize argon, ground
state configuration [Ne]$3s^23p^6~(l_0=1)$, dominant orbital-angular-momentum quantum
numbers of $l=5,6,5$ were found. These values for $l$ agree with those obtained for OCS, ground
state configuration $\ldots9\sigma^2{}3\pi^4~(l_0=1)$, providing clear evidence of the
empirical rule and its extension to molecules.

The time-dependent Schrödinger equation for the strong-field ionization of H and Ar pictured the
trajectory of the outgoing electron as a quiver motion along the laser-polarization axis
superimposed to a drift motion following a Kepler hyperbola~\cite{Arbo:PRA78:1418}. Identifying the
hyperbola's pericenter as the electron's quiver amplitude in the field provided an analytical
expression for an effective final orbital momentum that -- for single ionization -- only depends on
the electron's final linear momentum and its quiver amplitude in the field. For the ionization of
OCS through $10,11,12$ photons effective orbital quantum numbers $\tilde{l}=5.5,6.5,7.1$ were
obtained accordingly~\cite{Arbo:PRA78:1418}. Although this model considers neither the initial
state's orbital momentum nor a parity selection rule, the values for $\tilde{l}$ reflect our
experimental observations remarkably well. This further supports the statement that the momentum
distribution of an electron wave packet from non-resonant strong-field ionization is dictated by
the electric field and its Coulomb interaction with the cation. The initial bound electronic state
imprints onto the free-electron wave packet through its vertical ionization energy, its orbital
angular momentum, and the symmetry of its electron density distribution. While the impact of the
former two properties was demonstrated within the scope of this manuscript, the investigation of
the latter would require referencing to the molecular frame.

\section{Conclusions}
\label{sec:conclusions}
Three-dimensional photoelectron momentum distributions from strong-field ionization of OCS at
differential peak intensities were recorded and analyzed in matters of resonant Rydberg states and
photoelectron angular momentum. The evaluation of the derivatives of the momentum distributions with
respect to the peak intensity allowed for the unambiguous recognition of Freeman resonances. As a
result, sharp photoelectron lines, previously ascribed to valence-valence
transitions~\cite{Yu:JPB50:235602}, were reassigned to progressions of Freeman resonances
corresponding to Rydberg states with principal quantum numbers $n=4,5,7,8,9,10$. Furthermore, the
empirical rule that relates the initial state's angular momentum and the minimum photon expense to
ionize an ac Stark shifted atomic system to the observable dominant photoelectron
momentum~\cite{Chen:PRA74:053405} could be confirmed for a molecular target. In addition, the clear
closing of ionization channels with increasing peak intensity is shown for the OCS molecule, which
could eventually enable the probing of a distinct strong-field ionization channel in a
coherent-control experimental setup. It is demonstrated that in order to gain insight into
strong-field ionization processes on a quantum-mechanical level it is essential to relate distinct
incident intensities to the resulting response of the system under investigation. Especially for the
exploration of the transition regime between multi-photon and tunneling ionization a well-defined
and narrow intensity distribution is crucial. Studies employing a clear relation between incident
intensity and target response could finally shed light onto the importance of resonant states and,
thus, show the shortcomings of present molecular strong-field ionization theories to describe the
outgoing electron wave packet.

\section*{Acknowledgments}
This work has been supported by the Clusters of Excellence ``Center for Ultrafast Imaging'' (CUI,
EXC~1074, ID~194651731) and ``Advanced Imaging of Matter'' (AIM, EXC~2056, ID~390715994) of the
Deutsche Forschungsgemeinschaft (DFG), by the European Research Council under the European Union's
Seventh Framework Program (FP7/2007-2013) through the Consolidator Grant COMOTION
(ERC-Küpper-614507), and by the Helmholtz Gemeinschaft through the ``Impuls- und Vernetzungsfond''.
A.T.\ gratefully acknowledges a fellowship by the Alexander von Humboldt Foundation.

\bibliography{string,cmi}

\begin{thebibliography}{33}%
\makeatletter
\providecommand \@ifxundefined [1]{%
 \@ifx{#1\undefined}
}%
\providecommand \@ifnum [1]{%
 \ifnum #1\expandafter \@firstoftwo
 \else \expandafter \@secondoftwo
 \fi
}%
\providecommand \@ifx [1]{%
 \ifx #1\expandafter \@firstoftwo
 \else \expandafter \@secondoftwo
 \fi
}%
\providecommand \natexlab [1]{#1}%
\providecommand \enquote  [1]{``#1''}%
\providecommand \bibnamefont  [1]{#1}%
\providecommand \bibfnamefont [1]{#1}%
\providecommand \citenamefont [1]{#1}%
\providecommand \href@noop [0]{\@secondoftwo}%
\providecommand \href [0]{\begingroup \@sanitize@url \@href}%
\providecommand \@href[1]{\@@startlink{#1}\@@href}%
\providecommand \@@href[1]{\endgroup#1\@@endlink}%
\providecommand \@sanitize@url [0]{\catcode `\\12\catcode `\$12\catcode
  `\&12\catcode `\#12\catcode `\^12\catcode `\_12\catcode `\%12\relax}%
\providecommand \@@startlink[1]{}%
\providecommand \@@endlink[0]{}%
\providecommand \url  [0]{\begingroup\@sanitize@url \@url }%
\providecommand \@url [1]{\endgroup\@href {#1}{\urlprefix }}%
\providecommand \urlprefix  [0]{URL }%
\providecommand \Eprint [0]{\href }%
\providecommand \doibase [0]{http://dx.doi.org/}%
\providecommand \selectlanguage [0]{\@gobble}%
\providecommand \bibinfo  [0]{\@secondoftwo}%
\providecommand \bibfield  [0]{\@secondoftwo}%
\providecommand \translation [1]{[#1]}%
\providecommand \BibitemOpen [0]{}%
\providecommand \bibitemStop [0]{}%
\providecommand \bibitemNoStop [0]{.\EOS\space}%
\providecommand \EOS [0]{\spacefactor3000\relax}%
\providecommand \BibitemShut  [1]{\csname bibitem#1\endcsname}%
\let\auto@bib@innerbib\@empty
\bibitem [{\citenamefont {Hankin}\ \emph {et~al.}(2001)\citenamefont {Hankin},
  \citenamefont {Villeneuve}, \citenamefont {Corkum},\ and\ \citenamefont
  {Rayner}}]{Hankin:PRA64:013405}%
  \BibitemOpen
  \bibfield  {author} {\bibinfo {author} {\bibfnamefont {S.}~\bibnamefont
  {Hankin}}, \bibinfo {author} {\bibfnamefont {D.}~\bibnamefont {Villeneuve}},
  \bibinfo {author} {\bibfnamefont {P.}~\bibnamefont {Corkum}}, \ and\ \bibinfo
  {author} {\bibfnamefont {D.}~\bibnamefont {Rayner}},\ }\bibfield  {title}
  {\enquote {\bibinfo {title} {Intense-field laser ionization rates in atoms
  and molecules},}\ }\href@noop {} {\bibfield  {journal} {\bibinfo  {journal}
  {Phys.\ Rev.\ A}\ }\textbf {\bibinfo {volume} {64}},\ \bibinfo {pages}
  {013405} (\bibinfo {year} {2001})}\BibitemShut {NoStop}%
\bibitem [{\citenamefont {Hart}\ \emph {et~al.}(2014)\citenamefont {Hart},
  \citenamefont {Strohaber}, \citenamefont {Kaya}, \citenamefont {Kaya},
  \citenamefont {Kolomenskii},\ and\ \citenamefont
  {Schuessler}}]{Hart:PRA89:1393}%
  \BibitemOpen
  \bibfield  {author} {\bibinfo {author} {\bibfnamefont {N.~A.}\ \bibnamefont
  {Hart}}, \bibinfo {author} {\bibfnamefont {J.}~\bibnamefont {Strohaber}},
  \bibinfo {author} {\bibfnamefont {G.}~\bibnamefont {Kaya}}, \bibinfo {author}
  {\bibfnamefont {N.}~\bibnamefont {Kaya}}, \bibinfo {author} {\bibfnamefont
  {A.~A.}\ \bibnamefont {Kolomenskii}}, \ and\ \bibinfo {author} {\bibfnamefont
  {H.~A.}\ \bibnamefont {Schuessler}},\ }\bibfield  {title} {\enquote {\bibinfo
  {title} {Intensity-resolved above-threshold ionization of xenon with short
  laser pulses},}\ }\href {\doibase 10.1103/PhysRevA.89.053414} {\bibfield
  {journal} {\bibinfo  {journal} {Phys.\ Rev.\ A}\ }\textbf {\bibinfo {volume}
  {89}},\ \bibinfo {pages} {1393} (\bibinfo {year} {2014})}\BibitemShut
  {NoStop}%
\bibitem [{\citenamefont {Shao}\ \emph {et~al.}(2015)\citenamefont {Shao},
  \citenamefont {Li}, \citenamefont {Liu}, \citenamefont {Sun}, \citenamefont
  {Xie}, \citenamefont {Wang}, \citenamefont {Deng}, \citenamefont {Wu},
  \citenamefont {Gong},\ and\ \citenamefont {Liu}}]{Shao:PRA:2008}%
  \BibitemOpen
  \bibfield  {author} {\bibinfo {author} {\bibfnamefont {Y.}~\bibnamefont
  {Shao}}, \bibinfo {author} {\bibfnamefont {M.}~\bibnamefont {Li}}, \bibinfo
  {author} {\bibfnamefont {M.~M.}\ \bibnamefont {Liu}}, \bibinfo {author}
  {\bibfnamefont {X.}~\bibnamefont {Sun}}, \bibinfo {author} {\bibfnamefont
  {X.}~\bibnamefont {Xie}}, \bibinfo {author} {\bibfnamefont {P.}~\bibnamefont
  {Wang}}, \bibinfo {author} {\bibfnamefont {Y.}~\bibnamefont {Deng}}, \bibinfo
  {author} {\bibfnamefont {C.}~\bibnamefont {Wu}}, \bibinfo {author}
  {\bibfnamefont {Q.}~\bibnamefont {Gong}}, \ and\ \bibinfo {author}
  {\bibfnamefont {Y.}~\bibnamefont {Liu}},\ }\bibfield  {title} {\enquote
  {\bibinfo {title} {{Isolating resonant excitation from above-threshold
  ionization}},}\ }\href {\doibase 10.1103/PhysRevA.92.013415} {\bibfield
  {journal} {\bibinfo  {journal} {Phys.\ Rev.\ A}\ }\textbf {\bibinfo {volume}
  {92}},\ \bibinfo {pages} {2008} (\bibinfo {year} {2015})}\BibitemShut
  {NoStop}%
\bibitem [{\citenamefont {Yu}\ \emph {et~al.}(2017)\citenamefont {Yu},
  \citenamefont {Hu}, \citenamefont {Li}, \citenamefont {Ma}, \citenamefont
  {He}, \citenamefont {Liu}, \citenamefont {Wang}, \citenamefont {Luo},\ and\
  \citenamefont {Ding}}]{Yu:JPB50:235602}%
  \BibitemOpen
  \bibfield  {author} {\bibinfo {author} {\bibfnamefont {J.}~\bibnamefont
  {Yu}}, \bibinfo {author} {\bibfnamefont {W.}~\bibnamefont {Hu}}, \bibinfo
  {author} {\bibfnamefont {X.}~\bibnamefont {Li}}, \bibinfo {author}
  {\bibfnamefont {P.}~\bibnamefont {Ma}}, \bibinfo {author} {\bibfnamefont
  {L.}~\bibnamefont {He}}, \bibinfo {author} {\bibfnamefont {F.}~\bibnamefont
  {Liu}}, \bibinfo {author} {\bibfnamefont {C.}~\bibnamefont {Wang}}, \bibinfo
  {author} {\bibfnamefont {S.}~\bibnamefont {Luo}}, \ and\ \bibinfo {author}
  {\bibfnamefont {D.}~\bibnamefont {Ding}},\ }\bibfield  {title} {\enquote
  {\bibinfo {title} {{Contribution of resonance excitation on ionization of OCS
  molecules in strong laser fields}},}\ }\href {\doibase  10.1088/1361-6455/aa8e67} {\bibfield  {journal} {\bibinfo  {journal} {J.\
  Phys.\ B}\ }\textbf {\bibinfo {volume} {50}},\ \bibinfo {pages} {235602}
  (\bibinfo {year} {2017})}\BibitemShut {NoStop}%
\bibitem [{\citenamefont {Li}\ \emph {et~al.}(2015)\citenamefont {Li},
  \citenamefont {Zhang}, \citenamefont {Luo}, \citenamefont {Zhou},
  \citenamefont {Zhang}, \citenamefont {Lan},\ and\ \citenamefont
  {Lu}}]{Li:PRA92:1945}%
  \BibitemOpen
  \bibfield  {author} {\bibinfo {author} {\bibfnamefont {M.}~\bibnamefont
  {Li}}, \bibinfo {author} {\bibfnamefont {P.}~\bibnamefont {Zhang}}, \bibinfo
  {author} {\bibfnamefont {S.}~\bibnamefont {Luo}}, \bibinfo {author}
  {\bibfnamefont {Y.}~\bibnamefont {Zhou}}, \bibinfo {author} {\bibfnamefont
  {Q.}~\bibnamefont {Zhang}}, \bibinfo {author} {\bibfnamefont
  {P.}~\bibnamefont {Lan}}, \ and\ \bibinfo {author} {\bibfnamefont
  {P.}~\bibnamefont {Lu}},\ }\bibfield  {title} {\enquote {\bibinfo {title}
  {{Selective enhancement of resonant multiphoton ionization with strong laser
  fields}},}\ }\href {\doibase 10.1103/PhysRevA.92.063404} {\bibfield
  {journal} {\bibinfo  {journal} {Phys.\ Rev.\ A}\ }\textbf {\bibinfo {volume}
  {92}},\ \bibinfo {pages} {1945} (\bibinfo {year} {2015})}\BibitemShut
  {NoStop}%
\bibitem [{\citenamefont {Schyja}\ \emph {et~al.}(1998)\citenamefont {Schyja},
  \citenamefont {Lang},\ and\ \citenamefont {Helm}}]{Schyja:PRA57:3692}%
  \BibitemOpen
  \bibfield  {author} {\bibinfo {author} {\bibfnamefont {V.}~\bibnamefont
  {Schyja}}, \bibinfo {author} {\bibfnamefont {T.}~\bibnamefont {Lang}}, \ and\
  \bibinfo {author} {\bibfnamefont {H.}~\bibnamefont {Helm}},\ }\bibfield
  {title} {\enquote {\bibinfo {title} {{Channel switching in above-threshold
  ionization of xenon}},}\ }\href {\doibase 10.1103/PhysRevA.57.3692}
  {\bibfield  {journal} {\bibinfo  {journal} {Phys.\ Rev.\ A}\ }\textbf
  {\bibinfo {volume} {57}},\ \bibinfo {pages} {3692--3697} (\bibinfo {year}
  {1998})}\BibitemShut {NoStop}%
\bibitem [{\citenamefont {Marchenko}\ \emph {et~al.}(2010)\citenamefont
  {Marchenko}, \citenamefont {Muller}, \citenamefont {Schafer},\ and\
  \citenamefont {Vrakking}}]{Marchenko:JPB43:095601}%
  \BibitemOpen
  \bibfield  {author} {\bibinfo {author} {\bibfnamefont {T.}~\bibnamefont
  {Marchenko}}, \bibinfo {author} {\bibfnamefont {H.~G.}\ \bibnamefont
  {Muller}}, \bibinfo {author} {\bibfnamefont {K.~J.}\ \bibnamefont {Schafer}},
  \ and\ \bibinfo {author} {\bibfnamefont {M.~J.~J.}\ \bibnamefont
  {Vrakking}},\ }\bibfield  {title} {\enquote {\bibinfo {title} {{Electron
  angular distributions in near-threshold atomic ionization}},}\ }\href
  {\doibase 10.1088/0953-4075/43/9/095601} {\bibfield  {journal} {\bibinfo
  {journal} {J.\ Phys.\ B}\ }\textbf {\bibinfo {volume} {43}},\ \bibinfo
  {pages} {095601} (\bibinfo {year} {2010})}\BibitemShut {NoStop}%
\bibitem [{\citenamefont {Holmegaard}\ \emph {et~al.}(2010)\citenamefont
  {Holmegaard}, \citenamefont {Hansen}, \citenamefont {Kalh{\o}j},
  \citenamefont {Kragh}, \citenamefont {Stapelfeldt}, \citenamefont
  {Filsinger}, \citenamefont {K{\"u}pper}, \citenamefont {Meijer},
  \citenamefont {Dimitrovski}, \citenamefont {Abu-samha}, \citenamefont
  {Martiny},\ and\ \citenamefont {Madsen}}]{Holmegaard:NatPhys6:428}%
  \BibitemOpen
  \bibfield  {author} {\bibinfo {author} {\bibfnamefont {L.}~\bibnamefont
  {Holmegaard}}, \bibinfo {author} {\bibfnamefont {J.~L.}\ \bibnamefont
  {Hansen}}, \bibinfo {author} {\bibfnamefont {L.}~\bibnamefont {Kalh{\o}j}},
  \bibinfo {author} {\bibfnamefont {S.~L.}\ \bibnamefont {Kragh}}, \bibinfo
  {author} {\bibfnamefont {H.}~\bibnamefont {Stapelfeldt}}, \bibinfo {author}
  {\bibfnamefont {F.}~\bibnamefont {Filsinger}}, \bibinfo {author}
  {\bibfnamefont {J.}~\bibnamefont {K{\"u}pper}}, \bibinfo {author}
  {\bibfnamefont {G.}~\bibnamefont {Meijer}}, \bibinfo {author} {\bibfnamefont
  {D.}~\bibnamefont {Dimitrovski}}, \bibinfo {author} {\bibfnamefont
  {M.}~\bibnamefont {Abu-samha}}, \bibinfo {author} {\bibfnamefont {C.~P.~J.}\
  \bibnamefont {Martiny}}, \ and\ \bibinfo {author} {\bibfnamefont {L.~B.}\
  \bibnamefont {Madsen}},\ }\bibfield  {title} {\enquote {\bibinfo {title}
  {Photoelectron angular distributions from strong-field ionization of oriented
  molecules},}\ }\href {\doibase 10.1038/NPHYS1666} {\bibfield  {journal}
  {\bibinfo  {journal} {Nat. Phys.}\ }\textbf {\bibinfo {volume} {6}},\
  \bibinfo {pages} {428} (\bibinfo {year} {2010})},\ \Eprint
  {http://arxiv.org/abs/1003.4634} {arXiv:1003.4634 [physics]} \BibitemShut
  {NoStop}%
\bibitem [{\citenamefont {Dimitrovski}\ \emph {et~al.}(2015)\citenamefont
  {Dimitrovski}, \citenamefont {Maurer}, \citenamefont {Stapelfeldt},\ and\
  \citenamefont {Madsen}}]{Dimitrovski:JPB48:245601}%
  \BibitemOpen
  \bibfield  {author} {\bibinfo {author} {\bibfnamefont {D.}~\bibnamefont
  {Dimitrovski}}, \bibinfo {author} {\bibfnamefont {J.}~\bibnamefont {Maurer}},
  \bibinfo {author} {\bibfnamefont {H.}~\bibnamefont {Stapelfeldt}}, \ and\
  \bibinfo {author} {\bibfnamefont {L.~B.}\ \bibnamefont {Madsen}},\ }\bibfield
   {title} {\enquote {\bibinfo {title} {{Strong-field ionization of
  three-dimensionally aligned naphthalene molecules: orbital modification and
  imprints of orbital nodal planes}},}\ }\href {\doibase  10.1088/0953-4075/48/24/245601} {\bibfield  {journal} {\bibinfo  {journal}
  {J.\ Phys.\ B}\ }\textbf {\bibinfo {volume} {48}},\ \bibinfo {pages} {245601}
  (\bibinfo {year} {2015})}\BibitemShut {NoStop}%
\bibitem [{\citenamefont {Maurer}\ \emph {et~al.}(2012)\citenamefont {Maurer},
  \citenamefont {Dimitrovski}, \citenamefont {Christensen}, \citenamefont
  {Madsen},\ and\ \citenamefont {Stapelfeldt}}]{Maurer:PRL109:123001}%
  \BibitemOpen
  \bibfield  {author} {\bibinfo {author} {\bibfnamefont {J.}~\bibnamefont
  {Maurer}}, \bibinfo {author} {\bibfnamefont {D.}~\bibnamefont {Dimitrovski}},
  \bibinfo {author} {\bibfnamefont {L.}~\bibnamefont {Christensen}}, \bibinfo
  {author} {\bibfnamefont {L.~B.}\ \bibnamefont {Madsen}}, \ and\ \bibinfo
  {author} {\bibfnamefont {H.}~\bibnamefont {Stapelfeldt}},\ }\bibfield
  {title} {\enquote {\bibinfo {title} {Molecular-frame 3d photoelectron
  momentum distributions by tomographic reconstruction},}\ }\href {\doibase  10.1103/PhysRevLett.109.123001} {\bibfield  {journal} {\bibinfo  {journal}
  {Phys.\ Rev.\ Lett.}\ }\textbf {\bibinfo {volume} {109}},\ \bibinfo {pages}
  {123001} (\bibinfo {year} {2012})}\BibitemShut {NoStop}%
\bibitem [{\citenamefont {Ivanov}\ \emph {et~al.}(2005)\citenamefont {Ivanov},
  \citenamefont {Spanner},\ and\ \citenamefont
  {Smirnova}}]{Ivanov:JMODOPT52:165}%
  \BibitemOpen
  \bibfield  {author} {\bibinfo {author} {\bibfnamefont {M.~Y.}\ \bibnamefont
  {Ivanov}}, \bibinfo {author} {\bibfnamefont {M.}~\bibnamefont {Spanner}}, \
  and\ \bibinfo {author} {\bibfnamefont {O.}~\bibnamefont {Smirnova}},\
  }\bibfield  {title} {\enquote {\bibinfo {title} {Anatomy of strong field
  ionization},}\ }\href {\doibase 10.1080/0950034042000275360} {\bibfield
  {journal} {\bibinfo  {journal} {J.\ Mod.\ Opt.}\ }\textbf {\bibinfo {volume}
  {52}},\ \bibinfo {pages} {165--184} (\bibinfo {year} {2005})}\BibitemShut
  {NoStop}%
\bibitem [{\citenamefont {Ammosov}\ \emph {et~al.}(1986)\citenamefont
  {Ammosov}, \citenamefont {Delone},\ and\ \citenamefont
  {Krainov}}]{Ammosov:SVJETP64:1191}%
  \BibitemOpen
  \bibfield  {author} {\bibinfo {author} {\bibfnamefont {M.~V.}\ \bibnamefont
  {Ammosov}}, \bibinfo {author} {\bibfnamefont {N.~B.}\ \bibnamefont {Delone}},
  \ and\ \bibinfo {author} {\bibfnamefont {V.~P.}\ \bibnamefont {Krainov}},\
  }\bibfield  {title} {\enquote {\bibinfo {title} {Tunnel ionization of complex
  atoms and of atomic ions in an alternating electromagnetic field},}\ }\href
  {http://www.jetp.ac.ru/cgi-bin/e/index/e/64/6/p1191?a=list} {\bibfield
  {journal} {\bibinfo  {journal} {Soviet Physics - JETP}\ }\textbf {\bibinfo
  {volume} {64}},\ \bibinfo {pages} {1191--1194} (\bibinfo {year}
  {1986})}\BibitemShut {NoStop}%
\bibitem [{\citenamefont {Walker}\ \emph {et~al.}(1998)\citenamefont {Walker},
  \citenamefont {Hansch},\ and\ \citenamefont
  {Van~Woerkom}}]{Walker:PRA57:R701}%
  \BibitemOpen
  \bibfield  {author} {\bibinfo {author} {\bibfnamefont {M.~A.}\ \bibnamefont
  {Walker}}, \bibinfo {author} {\bibfnamefont {P.}~\bibnamefont {Hansch}}, \
  and\ \bibinfo {author} {\bibfnamefont {L.~D.}\ \bibnamefont {Van~Woerkom}},\
  }\bibfield  {title} {\enquote {\bibinfo {title} {{Intensity-resolved
  multiphoton ionization: Circumventing spatial averaging}},}\ }\href {\doibase  10.1103/PhysRevA.57.R701} {\bibfield  {journal} {\bibinfo  {journal} {Phys.\
  Rev.\ A}\ }\textbf {\bibinfo {volume} {57}},\ \bibinfo {pages} {R701--R704}
  (\bibinfo {year} {1998})}\BibitemShut {NoStop}%
\bibitem [{\citenamefont {Wang}\ \emph {et~al.}(2005)\citenamefont {Wang},
  \citenamefont {Sayler}, \citenamefont {Carnes}, \citenamefont {Esry},\ and\
  \citenamefont {Ben-Itzhak}}]{Wang:OL30:664}%
  \BibitemOpen
  \bibfield  {author} {\bibinfo {author} {\bibfnamefont {P.}~\bibnamefont
  {Wang}}, \bibinfo {author} {\bibfnamefont {A.~M.}\ \bibnamefont {Sayler}},
  \bibinfo {author} {\bibfnamefont {K.~D.}\ \bibnamefont {Carnes}}, \bibinfo
  {author} {\bibfnamefont {B.~D.}\ \bibnamefont {Esry}}, \ and\ \bibinfo
  {author} {\bibfnamefont {I.}~\bibnamefont {Ben-Itzhak}},\ }\bibfield  {title}
  {\enquote {\bibinfo {title} {{Disentangling the volume effect through
  intensity-difference spectra: application to laser-induced dissociation of
  H$_2^+$}},}\ }\href {\doibase 10.1364/OL.30.000664} {\bibfield  {journal}
  {\bibinfo  {journal} {Opt.\ Lett.}\ }\textbf {\bibinfo {volume} {30}},\
  \bibinfo {pages} {664} (\bibinfo {year} {2005})}\BibitemShut {NoStop}%
\bibitem [{\citenamefont {Strohaber}\ \emph {et~al.}(2010)\citenamefont
  {Strohaber}, \citenamefont {Kolomenskii},\ and\ \citenamefont
  {Schuessler}}]{Strohaber:PRA82:013403}%
  \BibitemOpen
  \bibfield  {author} {\bibinfo {author} {\bibfnamefont {J.}~\bibnamefont
  {Strohaber}}, \bibinfo {author} {\bibfnamefont {A.~A.}\ \bibnamefont
  {Kolomenskii}}, \ and\ \bibinfo {author} {\bibfnamefont {H.~A.}\ \bibnamefont
  {Schuessler}},\ }\bibfield  {title} {\enquote {\bibinfo {title}
  {Reconstruction of ionization probabilities from spatially averaged data in
  $n$ dimensions},}\ }\href {\doibase 10.1103/PhysRevA.82.013403} {\bibfield
  {journal} {\bibinfo  {journal} {Phys.\ Rev.\ A}\ }\textbf {\bibinfo {volume}
  {82}},\ \bibinfo {pages} {013403} (\bibinfo {year} {2010})}\BibitemShut
  {NoStop}%
\bibitem [{\citenamefont {Trippel}\ \emph {et~al.}(2013)\citenamefont
  {Trippel}, \citenamefont {Mullins}, \citenamefont {M{\"u}ller}, \citenamefont
  {Kienitz}, \citenamefont {D{\l}ugo{\l}\k{e}cki},\ and\ \citenamefont
  {K{\"u}pper}}]{Trippel:MP111:1738}%
  \BibitemOpen
  \bibfield  {author} {\bibinfo {author} {\bibfnamefont {S.}~\bibnamefont
  {Trippel}}, \bibinfo {author} {\bibfnamefont {T.}~\bibnamefont {Mullins}},
  \bibinfo {author} {\bibfnamefont {N.~L.~M.}\ \bibnamefont {M{\"u}ller}},
  \bibinfo {author} {\bibfnamefont {J.~S.}\ \bibnamefont {Kienitz}}, \bibinfo
  {author} {\bibfnamefont {K.}~\bibnamefont {D{\l}ugo{\l}\k{e}cki}}, \ and\
  \bibinfo {author} {\bibfnamefont {J.}~\bibnamefont {K{\"u}pper}},\ }\bibfield
   {title} {\enquote {\bibinfo {title} {Strongly aligned and oriented molecular
  samples at a {kHz} repetition rate},}\ }\href {\doibase  10.1080/00268976.2013.780334} {\bibfield  {journal} {\bibinfo  {journal}
  {Mol.\ Phys.}\ }\textbf {\bibinfo {volume} {111}},\ \bibinfo {pages} {1738}
  (\bibinfo {year} {2013})},\ \Eprint {http://arxiv.org/abs/1301.1826}
  {arXiv:1301.1826 [physics]} \BibitemShut {NoStop}%
\bibitem [{\citenamefont {Chang}\ \emph {et~al.}(2015)\citenamefont {Chang},
  \citenamefont {Horke}, \citenamefont {Trippel},\ and\ \citenamefont
  {K{\"u}pper}}]{Chang:IRPC34:557}%
  \BibitemOpen
  \bibfield  {author} {\bibinfo {author} {\bibfnamefont {Y.-P.}\ \bibnamefont
  {Chang}}, \bibinfo {author} {\bibfnamefont {D.~A.}\ \bibnamefont {Horke}},
  \bibinfo {author} {\bibfnamefont {S.}~\bibnamefont {Trippel}}, \ and\
  \bibinfo {author} {\bibfnamefont {J.}~\bibnamefont {K{\"u}pper}},\ }\bibfield
   {title} {\enquote {\bibinfo {title} {Spatially-controlled complex molecules
  and their applications},}\ }\href {\doibase 10.1080/0144235X.2015.1077838}
  {\bibfield  {journal} {\bibinfo  {journal} {Int.\ Rev.\ Phys.\ Chem.}\
  }\textbf {\bibinfo {volume} {34}},\ \bibinfo {pages} {557--590} (\bibinfo
  {year} {2015})},\ \Eprint {http://arxiv.org/abs/1505.05632} {arXiv:1505.05632
  [physics]} \BibitemShut {NoStop}%
\bibitem [{\citenamefont {Even}\ \emph {et~al.}(2000)\citenamefont {Even},
  \citenamefont {Jortner}, \citenamefont {Noy}, \citenamefont {Lavie},\ and\
  \citenamefont {Cossart-Magos}}]{Even:JCP112:8068}%
  \BibitemOpen
  \bibfield  {author} {\bibinfo {author} {\bibfnamefont {U.}~\bibnamefont
  {Even}}, \bibinfo {author} {\bibfnamefont {J.}~\bibnamefont {Jortner}},
  \bibinfo {author} {\bibfnamefont {D.}~\bibnamefont {Noy}}, \bibinfo {author}
  {\bibfnamefont {N.}~\bibnamefont {Lavie}}, \ and\ \bibinfo {author}
  {\bibfnamefont {N.}~\bibnamefont {Cossart-Magos}},\ }\bibfield  {title}
  {\enquote {\bibinfo {title} {Cooling of large molecules below 1~{K} and {H}e
  clusters formation},}\ }\href {\doibase 10.1063/1.481405} {\bibfield
  {journal} {\bibinfo  {journal} {J.\ Chem.\ Phys.}\ }\textbf {\bibinfo
  {volume} {112}},\ \bibinfo {pages} {8068--8071} (\bibinfo {year}
  {2000})}\BibitemShut {NoStop}%
\bibitem [{\citenamefont {Bordas}\ \emph {et~al.}(1996)\citenamefont {Bordas},
  \citenamefont {Paulig}, \citenamefont {Helm},\ and\ \citenamefont
  {Huestis}}]{Bordas:RSI67:2257}%
  \BibitemOpen
  \bibfield  {author} {\bibinfo {author} {\bibfnamefont {C.}~\bibnamefont
  {Bordas}}, \bibinfo {author} {\bibfnamefont {F.}~\bibnamefont {Paulig}},
  \bibinfo {author} {\bibfnamefont {H.}~\bibnamefont {Helm}}, \ and\ \bibinfo
  {author} {\bibfnamefont {D.~L.}\ \bibnamefont {Huestis}},\ }\bibfield
  {title} {\enquote {\bibinfo {title} {{Photoelectron imaging spectrometry:
  Principle and inversion method}},}\ }\href {\doibase 10.1063/1.1147044}
  {\bibfield  {journal} {\bibinfo  {journal} {Rev.\ Sci.\ Instrum.}\ }\textbf
  {\bibinfo {volume} {67}},\ \bibinfo {pages} {2257} (\bibinfo {year}
  {1996})}\BibitemShut {NoStop}%
\bibitem [{\citenamefont {Rallis}\ \emph {et~al.}(2014)\citenamefont {Rallis},
  \citenamefont {Burwitz}, \citenamefont {Andrews}, \citenamefont {Zohrabi},
  \citenamefont {Averin}, \citenamefont {De}, \citenamefont {Bergues},
  \citenamefont {Jochim}, \citenamefont {Voznyuk}, \citenamefont {Gregerson},
  \citenamefont {Gaire}, \citenamefont {Znakovskaya}, \citenamefont {McKenna},
  \citenamefont {Carnes}, \citenamefont {Kling}, \citenamefont {Ben-Itzhak},\
  and\ \citenamefont {Wells}}]{Rallis:RSI85:113105}%
  \BibitemOpen
  \bibfield  {author} {\bibinfo {author} {\bibfnamefont {C.~E.}\ \bibnamefont
  {Rallis}}, \bibinfo {author} {\bibfnamefont {T.~G.}\ \bibnamefont {Burwitz}},
  \bibinfo {author} {\bibfnamefont {P.~R.}\ \bibnamefont {Andrews}}, \bibinfo
  {author} {\bibfnamefont {M.}~\bibnamefont {Zohrabi}}, \bibinfo {author}
  {\bibfnamefont {R.}~\bibnamefont {Averin}}, \bibinfo {author} {\bibfnamefont
  {S.}~\bibnamefont {De}}, \bibinfo {author} {\bibfnamefont {B.}~\bibnamefont
  {Bergues}}, \bibinfo {author} {\bibfnamefont {B.}~\bibnamefont {Jochim}},
  \bibinfo {author} {\bibfnamefont {A.~V.}\ \bibnamefont {Voznyuk}}, \bibinfo
  {author} {\bibfnamefont {N.}~\bibnamefont {Gregerson}}, \bibinfo {author}
  {\bibfnamefont {B.}~\bibnamefont {Gaire}}, \bibinfo {author} {\bibfnamefont
  {I.}~\bibnamefont {Znakovskaya}}, \bibinfo {author} {\bibfnamefont
  {J.}~\bibnamefont {McKenna}}, \bibinfo {author} {\bibfnamefont {K.~D.}\
  \bibnamefont {Carnes}}, \bibinfo {author} {\bibfnamefont {M.~F.}\
  \bibnamefont {Kling}}, \bibinfo {author} {\bibfnamefont {I.}~\bibnamefont
  {Ben-Itzhak}}, \ and\ \bibinfo {author} {\bibfnamefont {E.}~\bibnamefont
  {Wells}},\ }\bibfield  {title} {\enquote {\bibinfo {title} {{Incorporating
  real time velocity map image reconstruction into closed-loop coherent
  control}},}\ }\href {\doibase 10.1063/1.4899267} {\bibfield  {journal}
  {\bibinfo  {journal} {Rev.\ Sci.\ Instrum.}\ }\textbf {\bibinfo {volume}
  {85}},\ \bibinfo {pages} {113105} (\bibinfo {year} {2014})}\BibitemShut
  {NoStop}%
\bibitem [{\citenamefont {Hickstein}\ \emph {et~al.}(2016)\citenamefont
  {Hickstein}, \citenamefont {Yurchak}, \citenamefont {Dhrubajyoti},
  \citenamefont {Shih},\ and\ \citenamefont {Gibson}}]{Hickstein:pyabel}%
  \BibitemOpen
  \bibfield  {author} {\bibinfo {author} {\bibfnamefont {D.~D.}\ \bibnamefont
  {Hickstein}}, \bibinfo {author} {\bibfnamefont {R.}~\bibnamefont {Yurchak}},
  \bibinfo {author} {\bibfnamefont {D.}~\bibnamefont {Dhrubajyoti}}, \bibinfo
  {author} {\bibfnamefont {C.-Y.}\ \bibnamefont {Shih}}, \ and\ \bibinfo
  {author} {\bibfnamefont {S.~T.}\ \bibnamefont {Gibson}},\ }\href {\doibase  10.5281/zenodo.47423} {\enquote {\bibinfo {title} {{PyAbel (v0.7): A Python
  Package for Abel Transforms}},}\ } (\bibinfo {year} {2016})\BibitemShut
  {NoStop}%
\bibitem [{\citenamefont {Freeman}\ \emph {et~al.}(1987)\citenamefont
  {Freeman}, \citenamefont {Bucksbaum}, \citenamefont {Milchberg},
  \citenamefont {Darack}, \citenamefont {Schumacher},\ and\ \citenamefont
  {Geusic}}]{Freeman:PRL59:1092}%
  \BibitemOpen
  \bibfield  {author} {\bibinfo {author} {\bibfnamefont {R.~R.}\ \bibnamefont
  {Freeman}}, \bibinfo {author} {\bibfnamefont {P.~H.}\ \bibnamefont
  {Bucksbaum}}, \bibinfo {author} {\bibfnamefont {H.}~\bibnamefont
  {Milchberg}}, \bibinfo {author} {\bibfnamefont {S.}~\bibnamefont {Darack}},
  \bibinfo {author} {\bibfnamefont {D.}~\bibnamefont {Schumacher}}, \ and\
  \bibinfo {author} {\bibfnamefont {M.~E.}\ \bibnamefont {Geusic}},\ }\bibfield
   {title} {\enquote {\bibinfo {title} {{Above-threshold ionization with
  subpicosecond laser pulses}},}\ }\href {\doibase 10.1103/PhysRevLett.59.1092}
  {\bibfield  {journal} {\bibinfo  {journal} {Phys.\ Rev.\ Lett.}\ }\textbf
  {\bibinfo {volume} {59}},\ \bibinfo {pages} {1092--1095} (\bibinfo {year}
  {1987})}\BibitemShut {NoStop}%
\bibitem [{\citenamefont {Korneev}\ \emph {et~al.}(2012)\citenamefont
  {Korneev}, \citenamefont {Popruzhenko}, \citenamefont {Goreslavski},
  \citenamefont {Yan}, \citenamefont {Bauer}, \citenamefont {Becker},
  \citenamefont {K{\"u}bel}, \citenamefont {Kling}, \citenamefont {R{\"o}del},
  \citenamefont {W{\"u}nsche},\ and\ \citenamefont
  {Paulus}}]{Korneev:PRL108:223601}%
  \BibitemOpen
  \bibfield  {author} {\bibinfo {author} {\bibfnamefont {P.~A.}\ \bibnamefont
  {Korneev}}, \bibinfo {author} {\bibfnamefont {S.~V.}\ \bibnamefont
  {Popruzhenko}}, \bibinfo {author} {\bibfnamefont {S.~P.}\ \bibnamefont
  {Goreslavski}}, \bibinfo {author} {\bibfnamefont {T.-M.}\ \bibnamefont
  {Yan}}, \bibinfo {author} {\bibfnamefont {D.}~\bibnamefont {Bauer}}, \bibinfo
  {author} {\bibfnamefont {W.}~\bibnamefont {Becker}}, \bibinfo {author}
  {\bibfnamefont {M.}~\bibnamefont {K{\"u}bel}}, \bibinfo {author}
  {\bibfnamefont {M.~F.}\ \bibnamefont {Kling}}, \bibinfo {author}
  {\bibfnamefont {C.}~\bibnamefont {R{\"o}del}}, \bibinfo {author}
  {\bibfnamefont {M.}~\bibnamefont {W{\"u}nsche}}, \ and\ \bibinfo {author}
  {\bibfnamefont {G.~G.}\ \bibnamefont {Paulus}},\ }\bibfield  {title}
  {\enquote {\bibinfo {title} {{Interference Carpets in Above-Threshold
  Ionization: From the Coulomb-Free to the Coulomb-Dominated Regime}},}\ }\href
  {\doibase 10.1103/PhysRevLett.108.223601} {\bibfield  {journal} {\bibinfo
  {journal} {Phys.\ Rev.\ Lett.}\ }\textbf {\bibinfo {volume} {108}},\ \bibinfo
  {pages} {223601} (\bibinfo {year} {2012})}\BibitemShut {NoStop}%
\bibitem [{\citenamefont {Freeman}\ and\ \citenamefont
  {Bucksbaum}(1991)}]{Freeman:JPB24:325}%
  \BibitemOpen
  \bibfield  {author} {\bibinfo {author} {\bibfnamefont {R.~R.}\ \bibnamefont
  {Freeman}}\ and\ \bibinfo {author} {\bibfnamefont {P.~H.}\ \bibnamefont
  {Bucksbaum}},\ }\bibfield  {title} {\enquote {\bibinfo {title}
  {{Investigations of above-threshold ionization using subpicosecond laser
  pulses}},}\ }\href {\doibase 10.1088/0953-4075/24/2/004} {\bibfield
  {journal} {\bibinfo  {journal} {J.\ Phys.\ B}\ }\textbf {\bibinfo {volume}
  {24}},\ \bibinfo {pages} {325--347} (\bibinfo {year} {1991})}\BibitemShut
  {NoStop}%
\bibitem [{\citenamefont {Linstrom}\ and\ \citenamefont
  {Mallard}(2017)}]{NIST:webbook:2017}%
  \BibitemOpen
  \bibinfo {editor} {\bibfnamefont {P.~J.}\ \bibnamefont {Linstrom}}\ and\
  \bibinfo {editor} {\bibfnamefont {W.~G.}\ \bibnamefont {Mallard}},\ eds.,\
  \href {\doibase 10.18434/T4D303} {\emph {\bibinfo {title} {{NIST} Chemistry
  WebBook, {NIST} {S}tandard {R}eference {D}atabase {N}umber 69}}}\ (\bibinfo
  {publisher} {National Institute of Standards and Technology},\ \bibinfo
  {address} {Gaithersburg MD, 20899},\ \bibinfo {year} {2017})\BibitemShut
  {NoStop}%
\bibitem [{\citenamefont {O'Brian}\ \emph {et~al.}(1994)\citenamefont
  {O'Brian}, \citenamefont {Kim}, \citenamefont {Lan}, \citenamefont
  {McIlrath},\ and\ \citenamefont {Lucatorto}}]{OBrian:PRA49:R649}%
  \BibitemOpen
  \bibfield  {author} {\bibinfo {author} {\bibfnamefont {T.~R.}\ \bibnamefont
  {O'Brian}}, \bibinfo {author} {\bibfnamefont {J.-B.}\ \bibnamefont {Kim}},
  \bibinfo {author} {\bibfnamefont {G.}~\bibnamefont {Lan}}, \bibinfo {author}
  {\bibfnamefont {T.~J.}\ \bibnamefont {McIlrath}}, \ and\ \bibinfo {author}
  {\bibfnamefont {T.~B.}\ \bibnamefont {Lucatorto}},\ }\bibfield  {title}
  {\enquote {\bibinfo {title} {{Verification of the ponderomotive approximation
  for the ac Stark shift in Xe Rydberg levels}},}\ }\href {\doibase  10.1103/PhysRevA.49.R649} {\bibfield  {journal} {\bibinfo  {journal} {Phys.\
  Rev.\ A}\ }\textbf {\bibinfo {volume} {49}},\ \bibinfo {pages} {R649--R652}
  (\bibinfo {year} {1994})}\BibitemShut {NoStop}%
\bibitem [{\citenamefont {Tanaka}\ \emph {et~al.}(1960)\citenamefont {Tanaka},
  \citenamefont {Jursa},\ and\ \citenamefont {LeBlanc}}]{Tanaka:JCP32:1205}%
  \BibitemOpen
  \bibfield  {author} {\bibinfo {author} {\bibfnamefont {Y.}~\bibnamefont
  {Tanaka}}, \bibinfo {author} {\bibfnamefont {A.~S.}\ \bibnamefont {Jursa}}, \
  and\ \bibinfo {author} {\bibfnamefont {F.~J.}\ \bibnamefont {LeBlanc}},\
  }\bibfield  {title} {\enquote {\bibinfo {title} {Higher ionization potentials
  of linear triatomic molecules. {II.} {CS$_2$}, {COS}, and {N$_2$O}},}\ }\href
  {\doibase 10.1063/1.1730875} {\bibfield  {journal} {\bibinfo  {journal} {J.\
  Chem.\ Phys.}\ }\textbf {\bibinfo {volume} {32}},\ \bibinfo {pages}
  {1205--1214} (\bibinfo {year} {1960})}\BibitemShut {NoStop}%
\bibitem [{\citenamefont {Sunanda}\ \emph {et~al.}(2012)\citenamefont
  {Sunanda}, \citenamefont {Rajasekhar}, \citenamefont {Saraswathy},\ and\
  \citenamefont {Jagatap}}]{Sunanda:JQuantSpec113:58}%
  \BibitemOpen
  \bibfield  {author} {\bibinfo {author} {\bibfnamefont {K.}~\bibnamefont
  {Sunanda}}, \bibinfo {author} {\bibfnamefont {B.~N.}\ \bibnamefont
  {Rajasekhar}}, \bibinfo {author} {\bibfnamefont {P.}~\bibnamefont
  {Saraswathy}}, \ and\ \bibinfo {author} {\bibfnamefont {B.~N.}\ \bibnamefont
  {Jagatap}},\ }\bibfield  {title} {\enquote {\bibinfo {title}
  {{Photo-absorption studies on carbonyl sulphide in 30,000--91,000 cm$^{-1}$
  region using synchrotron radiation}},}\ }\href {\doibase  10.1016/j.jqsrt.2011.09.009} {\bibfield  {journal} {\bibinfo  {journal} {J.\
  Quant.\ Spectrosc.\ Radiat.\ Transf.}\ }\textbf {\bibinfo {volume} {113}},\
  \bibinfo {pages} {58--66} (\bibinfo {year} {2012})}\BibitemShut {NoStop}%
\bibitem [{\citenamefont {Cossart-Magos}\ \emph {et~al.}(2003)\citenamefont
  {Cossart-Magos}, \citenamefont {Jungen}, \citenamefont {Xu},\ and\
  \citenamefont {Launay}}]{CossartMagos:JCP119:3219}%
  \BibitemOpen
  \bibfield  {author} {\bibinfo {author} {\bibfnamefont {C.}~\bibnamefont
  {Cossart-Magos}}, \bibinfo {author} {\bibfnamefont {M.}~\bibnamefont
  {Jungen}}, \bibinfo {author} {\bibfnamefont {R.}~\bibnamefont {Xu}}, \ and\
  \bibinfo {author} {\bibfnamefont {F.}~\bibnamefont {Launay}},\ }\bibfield
  {title} {\enquote {\bibinfo {title} {{High resolution absorption spectrum of
  jet-cooled OCS between 64 000 and 91 000 cm$^{-1}$}},}\ }\href {\doibase  10.1063/1.1587114} {\bibfield  {journal} {\bibinfo  {journal} {J.\ Chem.\
  Phys.}\ }\textbf {\bibinfo {volume} {119}},\ \bibinfo {pages} {3219--3233}
  (\bibinfo {year} {2003})}\BibitemShut {NoStop}%
\bibitem [{\citenamefont {Rudenko}\ \emph {et~al.}(2004)\citenamefont
  {Rudenko}, \citenamefont {Zrost}, \citenamefont {Schr{\"o}ter}, \citenamefont
  {De~Jesus}, \citenamefont {Feuerstein}, \citenamefont {Moshammer},\ and\
  \citenamefont {Ullrich}}]{Rudenko:JPB37:L407}%
  \BibitemOpen
  \bibfield  {author} {\bibinfo {author} {\bibfnamefont {A.}~\bibnamefont
  {Rudenko}}, \bibinfo {author} {\bibfnamefont {K.}~\bibnamefont {Zrost}},
  \bibinfo {author} {\bibfnamefont {C.~D.}\ \bibnamefont {Schr{\"o}ter}},
  \bibinfo {author} {\bibfnamefont {V.~L.~B.}\ \bibnamefont {De~Jesus}},
  \bibinfo {author} {\bibfnamefont {B.}~\bibnamefont {Feuerstein}}, \bibinfo
  {author} {\bibfnamefont {R.}~\bibnamefont {Moshammer}}, \ and\ \bibinfo
  {author} {\bibfnamefont {J.}~\bibnamefont {Ullrich}},\ }\bibfield  {title}
  {\enquote {\bibinfo {title} {Resonant structures in the low-energy electron
  continuum for single ionization of atoms in the tunnelling regime},}\ }\href
  {\doibase 10.1088/0953-4075/37/24/L03} {\bibfield  {journal} {\bibinfo
  {journal} {J.\ Phys.\ B}\ }\textbf {\bibinfo {volume} {37}},\ \bibinfo
  {pages} {L407--L413} (\bibinfo {year} {2004})}\BibitemShut {NoStop}%
\bibitem [{\citenamefont {Arb{\'o}}\ \emph {et~al.}(2006)\citenamefont
  {Arb{\'o}}, \citenamefont {Yoshida}, \citenamefont {Persson}, \citenamefont
  {Dimitriou},\ and\ \citenamefont {Burgd{\"o}rfer}}]{Arbo:PRL96:143003}%
  \BibitemOpen
  \bibfield  {author} {\bibinfo {author} {\bibfnamefont {D.~G.}\ \bibnamefont
  {Arb{\'o}}}, \bibinfo {author} {\bibfnamefont {S.}~\bibnamefont {Yoshida}},
  \bibinfo {author} {\bibfnamefont {E.}~\bibnamefont {Persson}}, \bibinfo
  {author} {\bibfnamefont {K.~I.}\ \bibnamefont {Dimitriou}}, \ and\ \bibinfo
  {author} {\bibfnamefont {J.}~\bibnamefont {Burgd{\"o}rfer}},\ }\bibfield
  {title} {\enquote {\bibinfo {title} {{Interference Oscillations in the
  Angular Distribution of Laser-Ionized Electrons near Ionization
  Threshold}},}\ }\href {\doibase 10.1103/PhysRevLett.96.143003} {\bibfield
  {journal} {\bibinfo  {journal} {Phys.\ Rev.\ Lett.}\ }\textbf {\bibinfo
  {volume} {96}},\ \bibinfo {pages} {143003} (\bibinfo {year}
  {2006})}\BibitemShut {NoStop}%
\bibitem [{\citenamefont {Chen}\ \emph {et~al.}(2006)\citenamefont {Chen},
  \citenamefont {Morishita}, \citenamefont {Le}, \citenamefont {Wickenhauser},
  \citenamefont {Tong},\ and\ \citenamefont {Lin}}]{Chen:PRA74:053405}%
  \BibitemOpen
  \bibfield  {author} {\bibinfo {author} {\bibfnamefont {Z.}~\bibnamefont
  {Chen}}, \bibinfo {author} {\bibfnamefont {T.}~\bibnamefont {Morishita}},
  \bibinfo {author} {\bibfnamefont {A.-T.}\ \bibnamefont {Le}}, \bibinfo
  {author} {\bibfnamefont {M.}~\bibnamefont {Wickenhauser}}, \bibinfo {author}
  {\bibfnamefont {X.~M.}\ \bibnamefont {Tong}}, \ and\ \bibinfo {author}
  {\bibfnamefont {C.~D.}\ \bibnamefont {Lin}},\ }\bibfield  {title} {\enquote
  {\bibinfo {title} {Analysis of two-dimensional photoelectron momentum spectra
  and the effect of the long-range coulomb potential in single ionization of
  atoms by intense lasers},}\ }\href {\doibase 10.1103/PhysRevA.74.053405}
  {\bibfield  {journal} {\bibinfo  {journal} {Phys.\ Rev.\ A}\ }\textbf
  {\bibinfo {volume} {74}},\ \bibinfo {pages} {053405} (\bibinfo {year}
  {2006})}\BibitemShut {NoStop}%
\bibitem [{\citenamefont {Arb{\'o}}\ \emph {et~al.}(2008)\citenamefont
  {Arb{\'o}}, \citenamefont {Dimitriou}, \citenamefont {Persson},\ and\
  \citenamefont {Burgd{\"o}rfer}}]{Arbo:PRA78:1418}%
  \BibitemOpen
  \bibfield  {author} {\bibinfo {author} {\bibfnamefont {D.~G.}\ \bibnamefont
  {Arb{\'o}}}, \bibinfo {author} {\bibfnamefont {K.~I.}\ \bibnamefont
  {Dimitriou}}, \bibinfo {author} {\bibfnamefont {E.}~\bibnamefont {Persson}},
  \ and\ \bibinfo {author} {\bibfnamefont {J.}~\bibnamefont {Burgd{\"o}rfer}},\
  }\bibfield  {title} {\enquote {\bibinfo {title} {{Sub-Poissonian angular
  momentum distribution near threshold in atomic ionization by short laser
  pulses}},}\ }\href {\doibase 10.1103/PhysRevA.78.013406} {\bibfield
  {journal} {\bibinfo  {journal} {Phys.\ Rev.\ A}\ }\textbf {\bibinfo {volume}
  {78}},\ \bibinfo {pages} {1418} (\bibinfo {year} {2008})}\BibitemShut
  {NoStop}%
\end{thebibliography}%
\end{document}